\documentclass[twocolumn,aps,pre,nofootinbib,showpacs]{revtex4} %Older version
\usepackage{graphicx}% Include figure files
\usepackage{dcolumn}% Align table columns on decimal point
\usepackage{bm}% bold math
\usepackage{amssymb}
\usepackage{pifont}
\usepackage{amsmath}
\usepackage{times}
\usepackage{color}
\usepackage{soul}

\newcommand{\be}{\begin{equation}}
\newcommand{\ee}{\end{equation}}
\newcommand{\ba}{\begin{align}}
\newcommand{\ea}{\end{align}}

\newcommand{\al}{\alpha_{_L}}
\newcommand{\ac}{\alpha_{_C}}
\newcommand{\ar}{\alpha_{_R}}
\newcommand{\ml}{m_{_L}}
\newcommand{\mc}{m_{_C}}
\newcommand{\mr}{m_{_R}}
\newcommand{\nl}{n_{_L}}
\newcommand{\nc}{n_{_C}}
\newcommand{\nr}{n_{_R}}
\newcommand{\To}{T_{_0}}
\newcommand{\DT}{\Delta T}
\newcommand{\Tl}{T_{_L}}
\newcommand{\Tr}{T_{_R}}
\newcommand{\gc}{\gamma_{c}}
\newcommand{\jif}{J_i^{(1)}}
\newcommand{\jis}{J_i^{(2)}}

\begin{document}
%\draft
%\wideabs{

\title{Thermal rectification in oscillator lattices with a ballistic spacer and next nearest-neighbor interactions}

\author{M.~Romero-Bastida and Jeanette-Ivonne~Amaya-Dur\'an}
\affiliation{SEPI ESIME-Culhuac\'an, Instituto Polit\'ecnico Nacional, 
Av. Santa Ana No. 1000, Col. San Francisco Culhuac\'an, Delegaci\'on Coyoacan, 
Distrito Federal 04430, Mexico}
\email{mromerob@ipn.mx}

\date{\today}

\begin{abstract}
In this work we study the asymmetric heat flow, i.e., thermal rectification, of a one-dimensional, mass-graded system consisting of a coupled harmonic oscillator lattice (ballistic spacer) and two diffusive leads attached to the boundaries of the former with both nearest-neighbor and next-nearest-neighbor (NNN) interactions. The latter enhance the rectification properties of the system and specially its independence on system size. The system presents a maximum rectification efficiency for a very precise value of the parameter that controls the coupling strength of the NNN interactions that depend on the temperature range wherein the device operates. The origin of this maximum value is the asymmetric local heat flow response corresponding to the NNN contribution at both sides of the lighter mass-loaded diffusive lead as quantified by the spectral properties. Upon variation of the system's parameters the performance of the device is always enhanced in the presence of NNN interactions.
\end{abstract}
%}

% 44.10.+i Heat conduction
% 05.60.-k Transport processes
% 05.45.-a Nonlinear dynamics and chaos
% 05.10.Gg Stochastic analysis methods

\maketitle

%%%%%%%%%%%%%%%%%%%%%%%%%%%%%%%%%%%%%%%%%%%%%%%%%%%%%%%%%%
\section{Introduction\label{sec:Intro}}
%%%%%%%%%%%%%%%%%%%%%%%%%%%%%%%%%%%%%%%%%%%%%%%%%%%%%%%%%%

In the past decades a renewed interest in heat flow control has been to a large extent driven by the intense study of the thermal rectification (TR) effect, which manifests itself as the asymmetry of heat current when the temperature difference is inverted. It is manifestly central to heat management and nanodevice design, and therefore has sprung a large body of studies aimed to find, improve, and even control it~\cite{Roberts11,Li12,Maldovan13}. After the first theoretical proposals involving hybrid structures of sequentially coupled segments of one-dimensional (1D) anharmonic lattices proposed to investigate TR~\cite{Terraneo02}, this effect has been widely investigated theoretically and later realized experimentally by means of an asymmetric nanotube structure~\cite{Chang06a}; generically, these TR devices are known as {\it thermal diodes}.

So far, previous work has determined the necessary conditions that will render TR in a given system. First, a symmetry breaking mechanism along the direction of the heat flux must be present. This symmetry breaking is most conveniently realized by merging two materials exhibiting different heat transport characteristics, which was indeed the strategy employed in the above mentioned theoretical proposal. Subsequently it was shown that a modified two-segment setup yields considerably improved rectification performance as compared to the original three-segment setup~\cite{Li04a,Hu05}. It is in these latter works that the second ingredient for TR was determined, namely the match or mismatch of spectral properties of the different parts afforded by the anharmonicity of the lattice. This property in turn implies a temperature and position-dependent conductivity~\cite{Peyrard06}.

There have been proposed various mechanisms for fulfilling the above mentioned general conditions to obtain TR~\cite{Eckmann06,Casati07,Leitner13,Liu14,Reid19}, but among them graded systems, i.e., inhomogeneous systems whose structure changes gradually in space, have been theoretically shown to be optimal materials for thermal diodes~\cite{Pereira10b,Pereira11,Wang12}. In particular, rectification properties of coupled nearest-neighbor (NN) anharmonic oscillator lattices have been extensively studied, specially those with a linear mass gradient along its length~\cite{Yang07}. TR has been shown to be robust for a wide value range of various structural parameters~\cite{Romero13}, to a change in the form of the mass distribution, to the inclusion of an onsite potential~\cite{Romero17a}, and can even be improved when next nearest-neighbor (NNN) interactions are considered~\cite{Romero17}.

However, one of the most severe and so far not completely solved problems of these devices is that their small rectification efficiency rapidly decays to zero as the system size increases~\cite{Hu06,Hu06a}. Therefore, searching for alternative rectifying systems to overcome the deterioration of the rectification effect for large systems is of great current interest. One of the first proposals consisted of a graded system with long-range interactions, such that all the oscillators with very different masses in the lattice interact with each other, leading to an increase of asymmetry which favor the TR and avoid its decay with system size~\cite{Pereira13,Chen15}. Recently, an alternative implementation of such heat rectifier has been suggested by means of linear chains of ions in trap lattices with trapping frequencies~\cite{Simon19}.

Within the context of oscillator models, an interesting proposal to address the TR decay with system size consists of a 1D segmented mass-graded harmonic oscillator lattice, with the boundary regions of the system (termed left and right lead) interacting with a substrate modeled by an onsite potential~\cite{Chen18}. It has been shown that the central segment, without interaction with an onsite potential and termed ballistic spacer, contributes crucially to remove dependence of rectification on the system size. This result seem to be quite robust upon variation of the system parameters and even to the presence of anharmonic interactions among the oscillators, which is particularly relevant considering possible experimental implementations as well as future technological applications.

Considering the promising results on improving the thermal rectification of a mass-graded anharmonic lattice by the addition of NNN interactions~\cite{Romero17} already mentioned, we reconsider the above proposed model in the presence of those same interactions. These have been previously shown to be relevant, within the context of oscillator models, in the study of energy localization~\cite{RomeroArias08} and in the study of the thermal conductivity for large system sizes~\cite{Xiong12,Xiong14}. Furthermore, the TR effect is optimized for a very precise value of the parameter that quantifies the strength of the contribution of the NNN interactions.

This paper is organized as follows: In Sec.~\ref{sec:Model} the model system and methodology are presented. Our results on the dependence of rectification on the strength of the NNN interactions and other involved parameters are reported in Sec.~\ref{sec:Res}. The discussion of the results, as well as our conclusions, are presented in Sec.~\ref{sec:Disc}.

%%%%%%%%%%%%%%%%%%%%%%%%%%%%%%%%%%%%%%%%%%%%%%%%%%%%%%%%%%%%%%%%%%%%%%%%
\section{system description\label{sec:Model}}
%%%%%%%%%%%%%%%%%%%%%%%%%%%%%%%%%%%%%%%%%%%%%%%%%%%%%%%%%%%%%%%%%%%%%%%%

\begin{figure}\centering
\includegraphics[width=0.75\linewidth,angle=0.0]{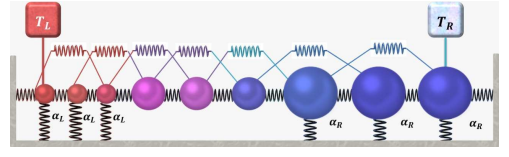}
\caption{(Color online) Schematic representation of a lattice with both NN and NNN interactions and connected at both ends to a thermal reservoir, each operating at different temperature. The central segment, composed of $\nc$ oscillators of mass $\mc$, is attached at the left (right) end to a lead composed of $\nl$ ($\nr$) oscillators of mass $\ml$ ($\mr$), which in turn interact with a substrate, modeled by an onsite potential of strength $\al$ ($\ar$).}
\label{fig:uno}
\end{figure}

The herein considered system, which is schematically depicted in Fig.~\ref{fig:uno}, is a 1D lattice of $N$ oscillators coupled both by a NN and NNN harmonic potential $V(x)= k_{_0}x^2/2$, where $k_{_0}=1$ is the harmonic constant. Furthermore, $\nl$ ($\nr$) oscillators with mass $\ml$ ($\mr$) in the left (right) side of the system are being acted upon by a quartic, $\phi^4$ onsite potential $U_{_{L,R}}(x)=\alpha_{_{L,R}}x^4/4$, whose strength is quantified by the magnitude of constant $\al$ ($\ar$). Therefore, the two anharmonic leads are connected by a purely harmonic lattice, i.e., a ballistic channel which would correspond to an onsite potential strength of $\ac=0$, composed of $\nc$ oscillators of mass $\mc$; thus, the total system size can be written as $N=\nl + \nc + \nr$. Fixed values of $\nl=\nr=10$ will be henceforth considered. Then the equations of motion (EOM) for each lattice oscillator can be written as $\dot q_i =p_i/m_i$ and
\begin{align}
\dot p_i & = F(q_i-q_{i-1})-F(q_{i+1}-q_i) \nonumber \\
   & + \gamma[F(q_i-q_{i-2})-F(q_{i+2}-q_i)] \nonumber \\
   & - \sum_{j=1}^{n_{_L}}\alpha_{_L}q_i^3\delta_{ij} - \sum_{k=N-n_{_R}+1}^N\alpha_{_R}q_i^3\delta_{ik} \nonumber \\
   & + (\xi_{_1} - \lambda_{_L} p_{_1})\,\delta_{1i} + (\xi_{_N} - \lambda_{_R} p_{_N})\,\delta_{Ni},
\label{EOM}
\end{align}
where $\{m_i,q_i,p_i\}_{i=1}^N$ are the dimensionless mass, displacement, and momentum of the $i$th oscillator; fixed boundary conditions are assumed ($q_{_0}=q_{_{N+1}}=0$). $F(x)=-\partial_x V(x)$ is the harmonic inter-oscillator force and the tunable parameter $\gamma$ specifies the relative strength of NNN coupling compared to the NN one. $\xi_{_{1,N}}$ is a Gaussian white noise with zero mean and correlation $\langle\xi_{_{1,N}}(t)\xi_{_{1,N}}(t^{\prime})\rangle=2\lambda_{_{1,N}}k_{_B}T_{_{1,N}}m_i(\delta_{1i}+\delta_{Ni})\delta(t-t^{\prime})$, with $\lambda_{_{1,N}}$ (taken as $=0.5$ in all computations hereafter reported) being the coupling strength between the first (last) oscillator in the lattice and the left (right) thermal reservoir operating at temperature $T_{_L}$ ($T_{_R}$). Therefore we can define the average temperature $\To\equiv(\Tl+\Tr)/2$ and difference $\DT\equiv\Tl-\Tr$; thus, $T_{_{L,R}}=\To\pm\DT/2$. In the following we will consider mass values as $\ml<\mc<\mr$, which amounts to a discontinuous right-to-left mass gradient. The above EOM~(\ref{EOM}) were integrated with a stochastic velocity-Verlet integrator with a time step of $\Delta t=10^{-2}$.

%%%%%%%%%%%%%%%%%%%%%%%%%%%%%%%%FIG. 1%%%%%%%%%%%%%%%%%%%%%%%%%%%%%%%%%%%%%%%%%%%%%%
%%%%%%%%%%%%%%%%%%%%%%%%%%%%%%%%FIG. 1%%%%%%%%%%%%%%%%%%%%%%%%%%%%%%%%%%%%%%%%%%%%%%

Equilibration times of $6\times10^7$ time units were needed to attain the nonequilibrium stationary state. The correctness of this computation was corroborated by comparing that the results of different equilibration times were equal; afterwards, the local heat flux is computed, just as in Ref.~\cite{Romero17}, as
\begin{align}
J_i& =\langle\dot q_{i} F(q_{i+1}-q_i)\rangle + 2\gamma\langle\dot q_{i+2} F(q_{i+2}-q_i)\rangle \nonumber \\
   & =\jif + 2\gamma\jis \label{lhf},
\end{align}
with $i\in[2,N-1]$ and $\langle\cdots\rangle$ indicating a time average over an interval of $8\times10^{7}$ time units. In the stationary state the local heat flux $J_i$ becomes constant along the lattice apart form thermal fluctuations, i.e. $J_i\sim J$, where $J$ is the total heat flux. Nevertheless, for the smallest $N$ values employed finite-size fluctuations in $J_i$ are not entirely negligible, as our results in the next section make evident. Therefore, to improve the accuracy of $J$ its value is calculated as the algebraic average of ${J}_i$ over the system bulk, i.e., excluding the oscillators connected to the reservoirs. We use $J_{+}$ to denote the heat flux obtained when the high temperature reservoir is attached to the heavy mass end and $J_{-}$ when it is attached to the light mass end. The \emph{rectifying efficiency} $r$ can thus be computed, to compare with the results of Ref.~\cite{Chen18}, from the expression
\be
r={(|J_+|-J_-)\over J_-}\times 100\%.
\ee
In the following the behavior of this quantity will be studied for low and high values of the average temperature $\To$ as well as a function of various structural parameters of the system.

%%%%%%%%%%%%%%%%%%%%%%%%%%%%%%%%%%%%%%%%%%%%%%%%%%%%%%%%%%%%%%%%%%%
\section{Results\label{sec:Res}}
%%%%%%%%%%%%%%%%%%%%%%%%%%%%%%%%%%%%%%%%%%%%%%%%%%%%%%%%%%%%%%%%%%%

\subsection{Maximum rectification efficiency}

\begin{figure}
\centerline{\includegraphics*[width=80mm]{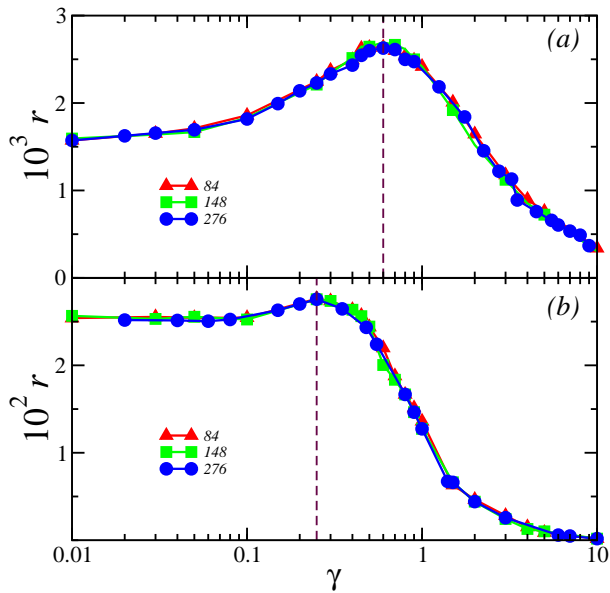}}
\caption{(Color online) (a) Thermal rectification $r$ vs. the relative strength of the NNN potential $\gamma$ for $N=84$ (circles), 148 (squares), and 276 (triangles) with $\ml=1$, $\mc=4.5$, $\mr=10$, $\al=\ar=1$, $\nl=\nr=10$, $\To=5$, and $\DT=9$. Vertical dotted line indicates the value $\gc=0.6$. (b) Same as in panel (a), but now for $\To=0.1$, and $\DT=0.16$. In this instance the vertical dashed line corresponds to $\gc=0.25$. Continuous lines are a guide to the eye.} 
\label{fig:dos}
\end{figure}

In Fig.~\ref{fig:dos}(a) we present the results of the dependence of the rectifying efficiency as a function of the relative strength $\gamma$ of the NNN potential for various system sizes $N$ and the same structural parameters as in the original study~\cite{Chen18}. First, it is important to notice the effect of the ballistic channel: the curves corresponding to various $N$ values overlap over most of the $\gamma$ range value, which is a clear signature that rectification is indeed system-size independent. Besides this new feature, the effect of the NNN interactions is very similar in this system as is on the linear mass-graded anharmonic system we previously studied~\cite{Romero17}: for small $\gamma$ values $r$ approaches the behavior of a lattice with only NN interactions, and in the opposite case, where NNN interactions are dominant, $r$ steadily declines, although rectification always remains significant, i.e., larger than the value corresponding to the aforementioned mass-graded lattice with NNN interactions previously considered. This behavior leads to the existence of a critical value, $\gc\simeq0.6$, where rectification is a maximum. Now, in Fig.~\ref{fig:dos}(b) we present the results corresponding to $\To=0.1$ and $\DT=0.16$, with similar results, except for two notable features. First, $r$, besides being lower in all the considered value range, is almost $\gamma$-independent for values $\gamma<0.1$. In this instance we have $\gc\simeq0.25$, lower than the value $\gc=0.45$ corresponding to the linear mass-graded anharmonic lattice~\cite{Romero17}. Therefore, the critical value is strongly dependent not only on the average temperature, but on the structural details of the considered lattice. Nevertheless, that $r$ is almost constant for $\gamma<0.1$ can be considered a desirable feature since, although the rectification value is lower than in the high-temperature instance, this independence of fine-tuned values of structural parameters could be relevant for future technological applications.

%%%%%%%%%%%%%%%%%%%%%%%%%%%%%%%%FIG. 2%%%%%%%%%%%%%%%%%%%%%%%%%%%%%%%%%%%%%%%%%%%%%%
%%%%%%%%%%%%%%%%%%%%%%%%%%%%%%%%FIG. 2%%%%%%%%%%%%%%%%%%%%%%%%%%%%%%%%%%%%%%%

\begin{figure}
\centerline{\includegraphics*[width=80mm]{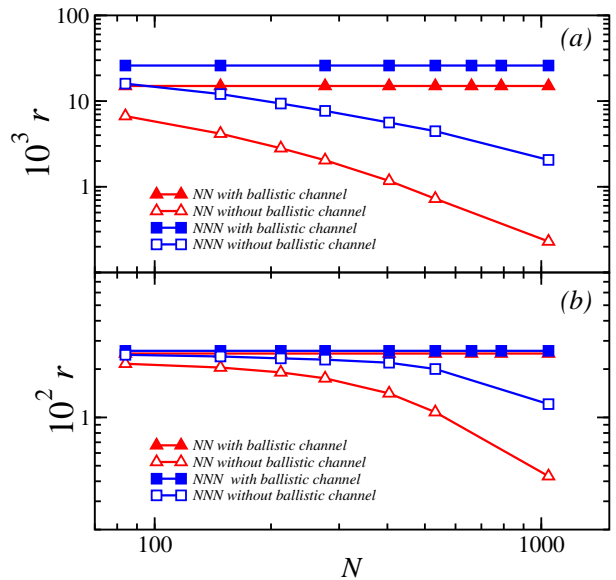}}
\caption{(Color online) (a) Thermal rectification $r$ vs. system size $N$ for lattices with NNN interactions (squares, $\gc=0.6$) and with NN interactions only (triangles). Filled symbols correspond to lattices with a ballistic channel and void symbols to lattices without one. Same parameter values as in Fig.~\ref{fig:dos}(a) with $\To=5$ and $\DT=9$. (b) Same as in panel (a), but now with $\To=0.1$, $\DT=0.16$, and $\gc=0.25$ for data corresponding to the lattice with NNN interactions.}
\label{fig:tres}
\end{figure}

In Fig.~\ref{fig:tres}(a) we plot the dependence of the rectification factor $r$ with system size $N$ in the high temperature case. For the lattice with no NNN interactions it is clear that for the case of a homogeneous onsite potential ($\al=\ac=\ar$) $r$ rapidly decays as the system size increases and, on the contrary, remains $N$-independent with the presence of the ballistic channel ($\ac=0$). The qualitative behavior is the same for the lattice with NNN interactions, with $r$ values higher for each $N$ value considered. However, there is a slight but important difference: for $\gamma=0$ (no NNN interactions) there is a $97\%$ decrease in $r$ when the system size changes from $N=84$ to $1044$, whereas the corresponding change for the $\gamma=0.6$ instance is of $87\%$. Thus, the NNN lattice performs better in the high $N$ limit. Now, for the low-temperature case the rectification for lattices with a ballistic channel is virtually the same, regardless of the presence of NNN interactions. But with lattices without the ballistic channel the decay in the absence of NNN interactions is more pronounced than in the case with them, since only for $N\ge500$ the decrease of $r$ becomes significant. Furthermore, the decrease for $\gamma=0$ is of $80\%$ and of $50\%$ for $\gamma=0.25$, which clearly indicates that the presence of NNN interactions diminishes the dependence of $r$ on $N$, just as in the case of mass-graded lattices~\cite{Romero17}, specially in the low-temperature regime. 

%%%%%%%%%%%%%%%%%%%%%%%%%%%%%%%%FIG. 3%%%%%%%%%%%%%%%%%%%%%%%%%%%%%%%%%%%
%%%%%%%%%%%%%%%%%%%%%%%%%%%%%%%%FIG. 3%%%%%%%%%%%%%%%%%%%%%%%%%%%%%%%%%%%%%%%

\subsection{Local heat flux and rectification}

\begin{figure}
\centerline{\includegraphics*[width=80mm]{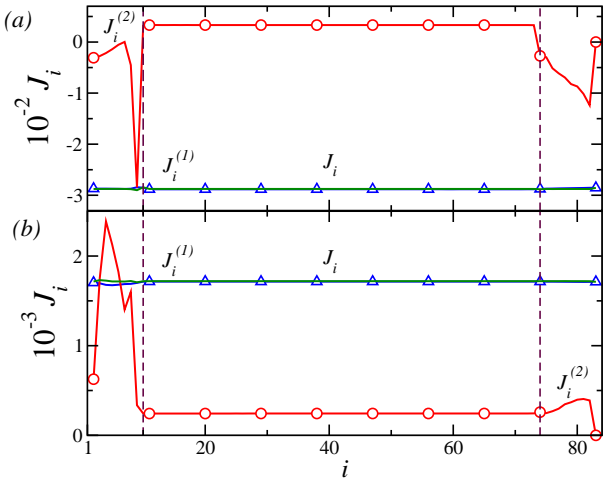}}
\caption{(Color online) Local heat flux for a lattice with $N=84$ and $\gamma=0.01$; same $\nl$, $\nr$, $\al$, and $\ar$ values as in Fig.~\ref{fig:dos}(a) with $\To=5$ and $\DT=9$. Lines with triangles indicate the NN contribution, lines with circles the NNN one, and solid lines the result according to Eq.~\ref{lhf} for (a) $\Tl<\Tr$ and (b) $\Tl>\Tr$. Vertical dashed lines indicate the boundaries with the leads.}
\label{fig:cuatro}
\end{figure}

In Fig.~\ref{fig:cuatro} we plot the contributions to the local heat flux along the lattice length from NN and NNN interactions, i.e., $\jif$ and $\jis$, for reverse (left-right) and forward (right-left) temperature bias in the high temperature regime, both with a $\gamma=0.01$ value. The latter corresponds to a setup wherein the relative contribution of NNN interactions is marginal. Therefore, when only the NN interactions are relevant, $r$ has a high value since not only $|J_+|>J_-$, but also there is an order-of-magnitude difference between both fluxes along the lattice length. However, at some sites in the left lead the magnitude of $\jis$ is close to that of the local heat flux $J_i$ for $\Tl<\Tr$ and is even bigger than $J_i$ when $\Tl>\Tr$. Since this peculiar behavior occurs in the case when the left lead is in contact with the hot reservoir ---that is, in a configuration wherewith the heat flux along the system is greatly diminished---, it seems reasonable to infer that the behavior of the term $\jis$ in that same lead might play an important role controlling the heat flux in the reverse temperature bias configuration for $\gamma>0.01$ values.

%%%%%%%%%%%%%%%%%%%%%%%%%%%%%%%%FIG. 4%%%%%%%%%%%%%%%%%%%%%%%%%%%%%%%%%%%%%%%%
%%%%%%%%%%%%%%%%%%%%%%%%%%%%%%%%FIG. 4%%%%%%%%%%%%%%%%%%%%%%%%%%%%%%%%%%%%%%%%

\begin{figure}
\centerline{\includegraphics*[width=60mm]{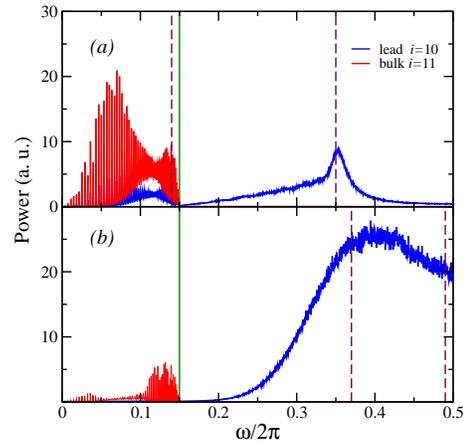}}
\caption{(Color online) (a) Power spectra for an oscillator in the left lead $i=10$ (blue) and one in the bulk $i=11$ (red) for $\gamma=0.01$ with $N=84$, $\To=5$, $\DT=9$, and $\Tl<\Tr$; same $\nl$, $\nr$, $\al$, and $\ar$ values as in Fig.~\ref{fig:dos}(a). (b) Same as panel (a) but for $\Tl>\Tr$. In both instances vertical dashed lines indicate the lower and upper limits of the lead phonon band; vertical solid line, upper limit of the bulk phonon band. See text for details.}
\label{fig:cinco}
\end{figure}

To study in more detail the origin of TR when the contribution of NN interactions to $J_i$ are predominant in Fig.~\ref{fig:cinco} we present the power spectra $|\tau^{-1}\!\!\int_{_0}^{\tau}\!\! dt\dot q_i(t)\exp(-\mathrm{i}\omega t)|^2$ of interface oscillators at the left (lead) and right (bulk) side of the boundary between both regions. For the $J_+$ configuration depicted in the upper panel the bulk spectrum lies within the low-frequency region and has a distinctly discrete structure characteristic of the underlying harmonic dynamics within the bulk. Thus, the phonon band is given by $[0,(4k_{_0}/\mc)^{\frac{1}{2}}]$; the upper limit, $\omega_{\mathrm{max}}/2\pi\sim0.15$, nicely coincides with the right boundary of the spectrum. On the other hand, in the lead region spectral power is more concentrated on the high-frequency region, a behavior consistent with the existence of the onsite anharmonic potential. For the latter the effective phonon approach~\cite{Li13} predicts that the active vibration frequencies are located within $[(1.23\Tl^{\frac{2}{3}})^{\frac{1}{2}},(4(k_{_0}/\ml)+1.23\Tl^{\frac{2}{3}})^{\frac{1}{2}}]$. It can be seen that the lower and upper limits of the phonon frequencies are in good agreement with the predicted phonon band $[0.14,0.35]$. However, due to the coupling between the lead and ballistic spacer ---the phonon band computed in~Ref.~\cite{Li13} is for a $\phi^4$ lattice that corresponds to the lead in the present study--- the phonon spectrum for the latter has an active low-frequency band that increases the frequency range wherein both spectra overlap. Thus, heat flux is favored in the right-left direction. Next, for the $J_-$ configuration depicted in Fig.~\ref{fig:cinco}(b) the contribution of the bulk spectrum is greatly diminished, whereas that of the lead is characterized by a strong activation of high-frequency phonons in the band $[0.37,0.49]$ predicted by the effective phonon approach. Therefore, the overlap with the bulk spectrum is almost suppressed, which in turn leads to an appreciable decrease in $J_-$ and a high rectification figure.

%%%%%%%%%%%%%%%%%%%%%%%%%%%%%%%%FIG. 5%%%%%%%%%%%%%%%%%%%%%%%%%%%%%%%%%%%%%%%%
%%%%%%%%%%%%%%%%%%%%%%%%%%%%%%%%FIG. 5%%%%%%%%%%%%%%%%%%%%%%%%%%%%%%%%%%%%%%%%

\begin{figure}
\centerline{\includegraphics*[width=80mm]{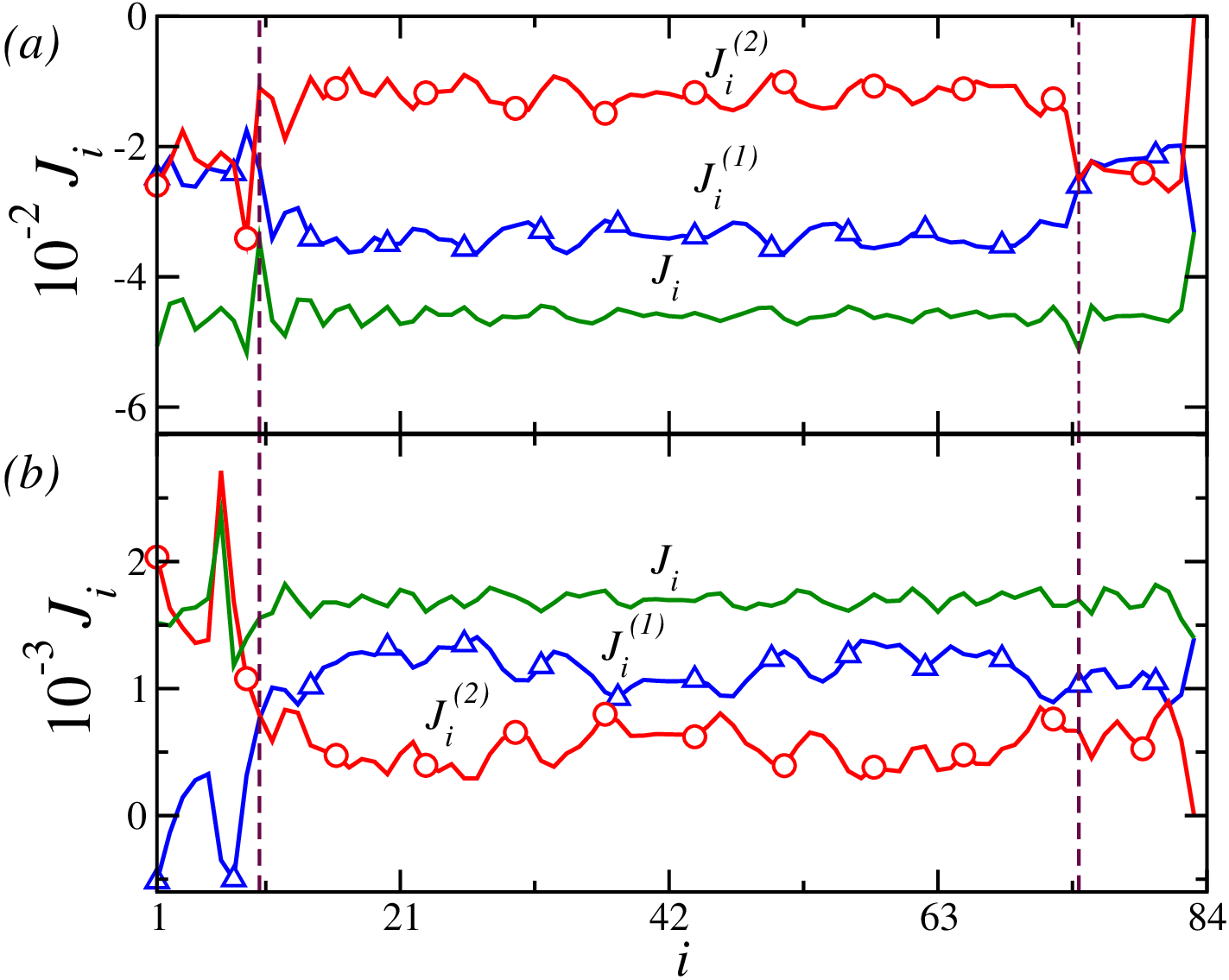}}
\caption{(Color online) Same as described in the caption of Fig.~\ref{fig:cuatro}, but for $\gc=0.6$ (a) $\Tl<\Tr$ and (b) $\Tl>\Tr$.} 
\label{fig:seis}
\end{figure}

Next, for the same temperature regime, in Fig.~\ref{fig:seis} we plot the local heat fluxes corresponding to $\gc=0.6$, which renders the maximum rectification figure displayed in Fig.~\ref{fig:dos}(a). It can be readily ascertained that now the contribution of the NNN interactions $\jis$ to the total value of $J_i$ becomes significant in the bulk region of the lattice for both temperature bias configurations. However, for the case of forward bias presented in Fig.~\ref{fig:seis}(a) the contributions of both NN and NNN interactions become of approximately the same magnitude in the boundary regions corresponding to the anharmonic leads. For reverse temperature bias, Fig.~\ref{fig:seis}(b), $\jif$ drastically drops in the left lead, connected to the hot reservoir in this configuration. Thus, the main contribution to local heat flux comes from the one corresponding to NNN interactions, i.e., $J_i\sim\jis$. Therefore, we can tentatively speculate that, in this case, the large rectification ---again, $J_+$ and $J_-$ differ by an order of magnitude as in the $\gamma=0.01$ case--- is due to an effect of the left lead on the NNN contribution to the heat flux when the former is in contact with the hot reservoir.

%%%%%%%%%%%%%%%%%%%%%%%%%%%%%%%%FIG. 6%%%%%%%%%%%%%%%%%%%%%%%%%%%%%%%%%%%%%%%%
%%%%%%%%%%%%%%%%%%%%%%%%%%%%%%%%FIG. 6%%%%%%%%%%%%%%%%%%%%%%%%%%%%%%%%%%%%%%%%

\begin{figure}
\centerline{\includegraphics*[width=60mm]{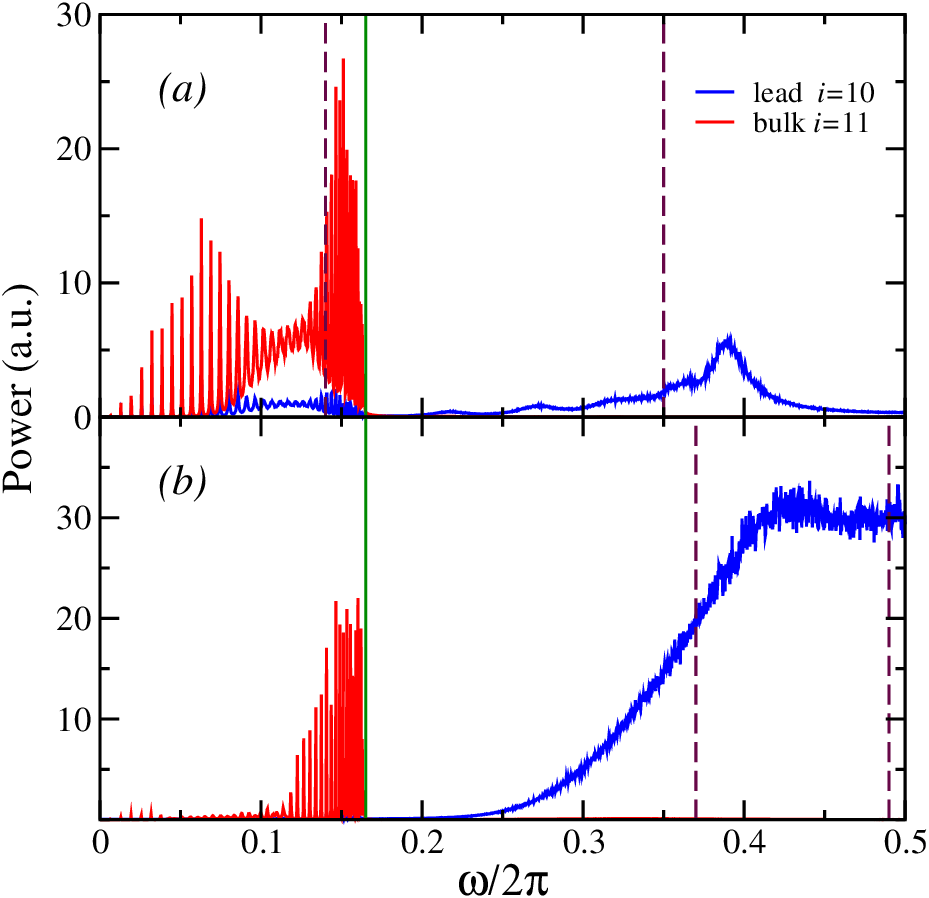}}
\caption{(Color online) Same as described in the caption of Fig.~\ref{fig:cinco}, but for $\gc=0.6$ (a) $\Tl<\Tr$ and (b) $\Tl>\Tr$.}
\label{fig:siete}
\end{figure}

The corresponding spectra for $\gc=0.6$ in the high-temperature regime are presented in Fig.~\ref{fig:siete}. In general, the phenomenology is similar to the case $\gamma=0.01$ previously studied, but with some caveats. For the forward-bias instance plotted in Fig.~\ref{fig:siete}(a) the bulk spectrum is shifted to higher frequencies. In this case the dispersion relation reads $\omega_{\alpha}=2[(k_{_0}/\ml)(\sin^2q_{\alpha}/2+\gamma\sin^2q_{\alpha})]^{\frac{1}{2}}$, where $q_{\alpha}$ is the wave number and $\omega_{\alpha}$ the corresponding frequency. For $\gamma=\gc$ we have $\omega_{\mathrm{max}}/2\pi\sim0.165$, which coincides with the high-frequency limit of the bulk spectrum. The lead spectrum is also shifted to high-frequencies, beyond the limits obtained from the effective phonon approach. This result is not unexpected because the currently available estimates of the effective phonon approach do not consider the contribution of NNN interactions~\cite{Li06a}. In the reverse-bias configuration, Fig.~\ref{fig:siete}(b), the lead spectrum is more concentrated around higher frequencies than the corresponding one for $\gamma=0.01$, which correlates well with what was observed in Fig.~\ref{fig:seis}(b), wherein the contribution of $\jis$ to $J_i$ is dominant in the lead region. Therefore, so far it seems that the left lead hinders the transmission of low-frequency, heat-carrying phonons associated with the NN contribution. Now, since the NNN contribution $\jis$ seems to be mainly composed of high-frequency phonons and $\jis<\jif$ in the bulk but $\jis\sim J_i$ in the lead, it is clear that in the latter only the less significant $\jis$ contribution remains, reducing $J_-$ and thus increasing $r$.

%%%%%%%%%%%%%%%%%%%%%%%%%%%%%%%%FIG. 7%%%%%%%%%%%%%%%%%%%%%%%%%%%%%%%%%%%%%%%%
%%%%%%%%%%%%%%%%%%%%%%%%%%%%%%%%FIG. 7%%%%%%%%%%%%%%%%%%%%%%%%%%%%%%%%%%%%%%%%

\begin{figure}
\centerline{\includegraphics*[width=80mm]{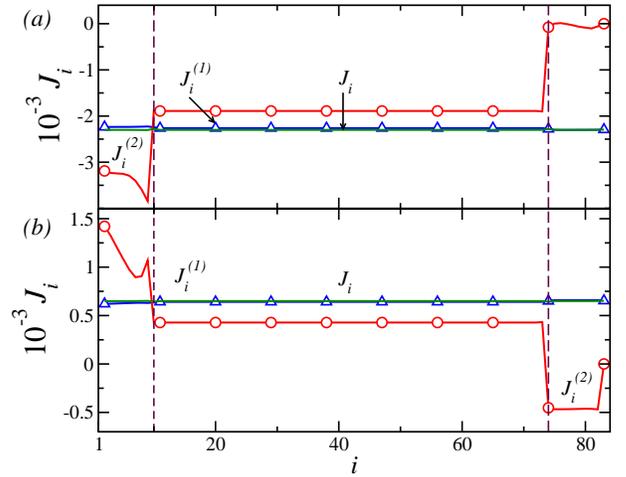}}
\caption{(Color online) Local heat flux for a lattice with $N=84$ and $\gamma=0.01$; same $\nl$, $\nr$, $\al$, and $\ar$ values as in Fig.~\ref{fig:dos}(b) with $\To=0.1$ and $\DT=0.16$. Lines with triangles indicate the NN contribution, lines with circles the NNN one, and solid lines the result according to Eq.~(\ref{lhf}) for (a) $\Tl<\Tr$ and (b) $\Tl>\Tr$. Vertical dashed lines indicate the boundaries with the leads.}
\label{fig:ocho}
\end{figure}

For the low-temperature case the results for local NN and NNN heat fluxes with $\gamma=0.01$ are presented in Fig.~\ref{fig:ocho}. Just as in the high-temperature case depicted in Fig.~\ref{fig:cuatro} the main contribution to the local heat flux is afforded by the NN contribution, i.e., $J_i\sim\jif$. The most important difference with the high-temperature instance is that the absolute value of $\jis$ is comparable to that of $\jif$ although, due to the low $\gamma=0.01$ value, its total contribution to $J_i$ is insignificant. Also when the applied temperature gradient is in the same direction as the mass gradient, see upper panel, $\jis$ increases ---discontinuously because of the segmented mass distribution--- along the system's length, but decreases when the temperature gradient is reversed, see lower panel. This latter behavior is different to that of the linear mass-graded anharmonic lattice studied in Ref.~\cite{Romero17}, wherein $\jis$ always increases in the direction where temperature decreases.

%%%%%%%%%%%%%%%%%%%%%%%%%%%%%%%%FIG. 8%%%%%%%%%%%%%%%%%%%%%%%%%%%%%%%%%%%%%%%%
%%%%%%%%%%%%%%%%%%%%%%%%%%%%%%%%FIG. 8%%%%%%%%%%%%%%%%%%%%%%%%%%%%%%%%%%%%%%%%

\begin{figure}
\centerline{\includegraphics*[width=60mm]{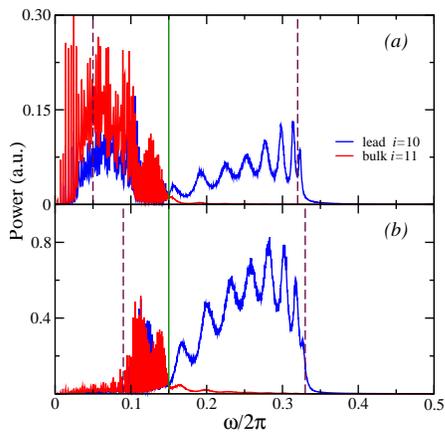}}
\caption{(Color online) (a) Power spectra for an oscillator in the left lead $i=10$ (blue) and one in the bulk $i=11$ (red) for $\gamma=0.01$ with $N=84$, $\To=0.1$, $\DT=0.16$, and $\Tl<\Tr$; same $\nl$, $\nr$, $\al$, and $\ar$ values as in Fig.~\ref{fig:dos}(a). (b) Same as panel (a) but for $\Tl>\Tr$.}
\label{fig:nueve}
\end{figure}

The corresponding spectra for the forward- and reverse-bias configurations with $\gamma=0.01$ and low average temperature are presented in Fig.~\ref{fig:nueve}. In the former case it is evident that the most significant contribution to both spectra lies in the low-frequency range, being this feature more extreme for the bulk spectrum. It is seen that, due to the reduced value of $\Tl=0.02$, the lead phonon band is broader, being $[0.05,0.32]$. This entails a large overlap with the bulk phonon band $[0,0.15]$. Thus, the low-frequency phonons that come from the bulk and are associated with the NN term $\jif$ are more easily transmitted through the left lead, which accounts for the preferred right-to-left heat flux direction. In the case of reverse-bias configuration depicted in Fig.~\ref{fig:nueve}(b) the lead spectrum, now with $\Tl=0.18$, is narrowed to $[0.09,0.33]$. The ensuing reduction in the overlap of both spectra reduces the magnitude of the heat flow in the reverse-bias configuration. In this temperature range the lead spectrum has a weaker temperature dependence, and thus the shift in the reverse-bias configuration is not as pronounced as that depicted in Figs.~\ref{fig:cinco} and~\ref{fig:siete}. Therefore, $J_+$ and $J_-$ are of the same order of magnitude and the ensuing $r$ value is also an order of magnitude lower than the corresponding one in the high-temperature instance, as was already observed in Fig.~\ref{fig:dos}.

%%%%%%%%%%%%%%%%%%%%%%%%%%%%%%%%FIG. 9%%%%%%%%%%%%%%%%%%%%%%%%%%%%%%%%%%%%%%%%
%%%%%%%%%%%%%%%%%%%%%%%%%%%%%%%%FIG. 9%%%%%%%%%%%%%%%%%%%%%%%%%%%%%%%%%%%%%%%%

\begin{figure}
\centerline{\includegraphics*[width=80mm]{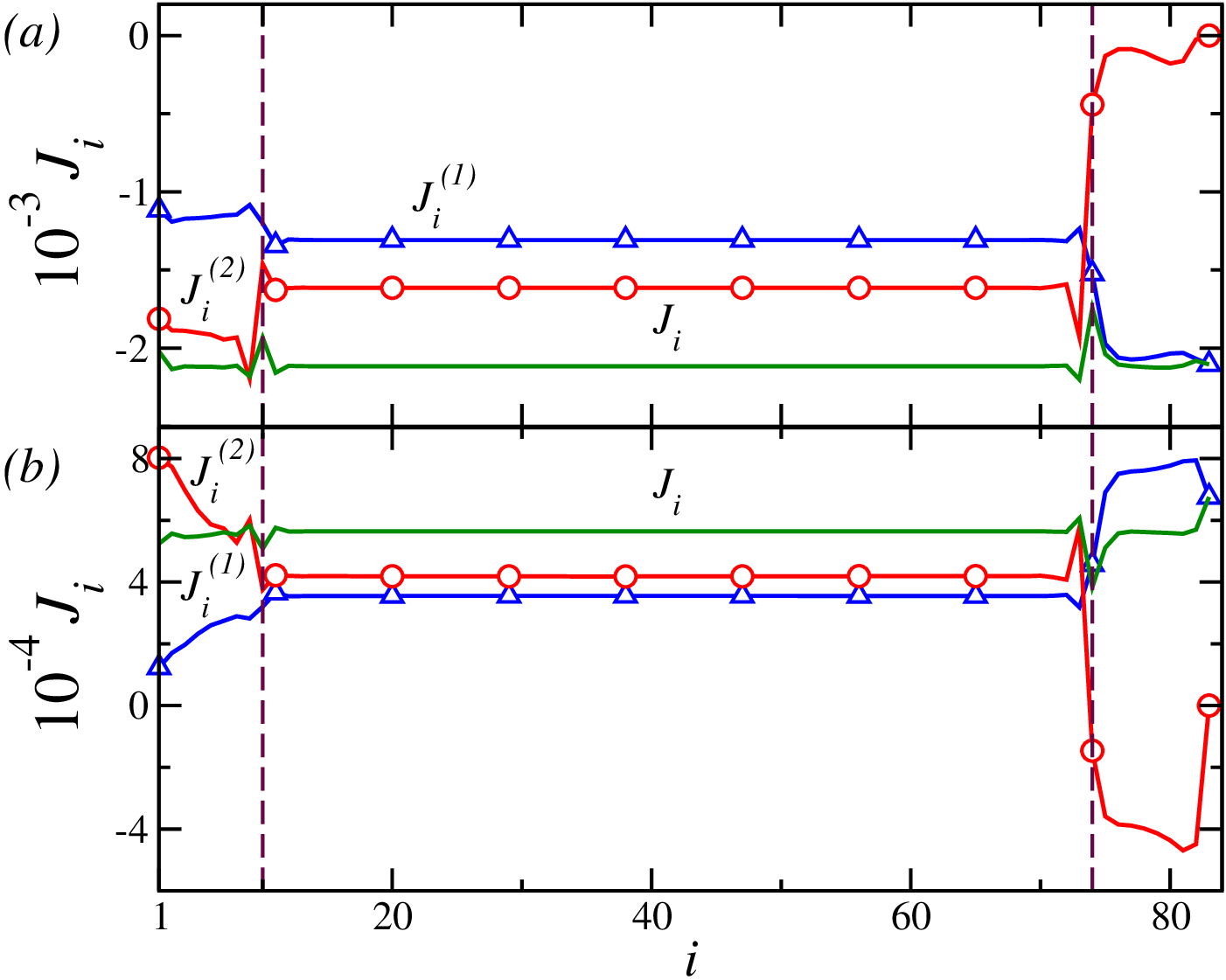}}
\caption{(Color online) Same as described in the caption of Fig.~\ref{fig:ocho}, but for $\gc=0.25$ (a) $\Tl<\Tr$ and (b) $\Tl>\Tr$.} 
\label{fig:diez}
\end{figure}

The NN and NNN terms of the local heat flux for which the $\gc$ value with which TR is maximized as observed in Fig.~\ref{fig:dos}(b) are presented in Fig.~\ref{fig:diez}. The same behavior of $\jis$ in the leads already noticed in Fig.~\ref{fig:ocho} is also present in this case; but now $\jis>\jif$ in the bulk region. This feature is common for both $J_+$ and $J_-$ configurations and is relevant to the heat flow along the system since now $\gamma=0.25$, and thus the $\jis$ contribution has to be more thoroughly considered. For the forward-bias case depicted in Fig.~\ref{fig:diez}(a) the difference between the magnitudes of $\jif$ and $\jis$ is greater inside the lead region than in the bulk. However, the change in magnitude of $\jif$ when going from the bulk to the left lead is less pronounced than that of $\jis$. Thus, the overall contribution to $J_i$ stems mainly from the $\jif$ term, just as in the $\gamma=0.01$ case depicted in Fig.~\ref{fig:ocho}, which favors the heat flow into the cold reservoir. The reverse-bias case presented in Fig.~\ref{fig:diez}(b) reveals that $J_i\sim\jis$ in the left lead, which is consistent with all previous cases for $J_-$ presented in Figs.~\ref{fig:seis} and~\ref{fig:ocho} wherein a high contribution of the NNN terms to the local heat flux leads to a reduction in $J_-$ magnitude and thus to high rectification. Furthermore, the same applies for $\gamma<\gc$ since there is a weak dependence of $r$ on $\gamma$ as was already observed in Fig.~\ref{fig:dos}(b). 

%%%%%%%%%%%%%%%%%%%%%%%%%%%%%%%%FIG. 10%%%%%%%%%%%%%%%%%%%%%%%%%%%%%%%%%%%%%%%
%%%%%%%%%%%%%%%%%%%%%%%%%%%%%%%%FIG. 10%%%%%%%%%%%%%%%%%%%%%%%%%%%%%%%%%%%%%%%

\begin{figure}
\centerline{\includegraphics*[width=60mm]{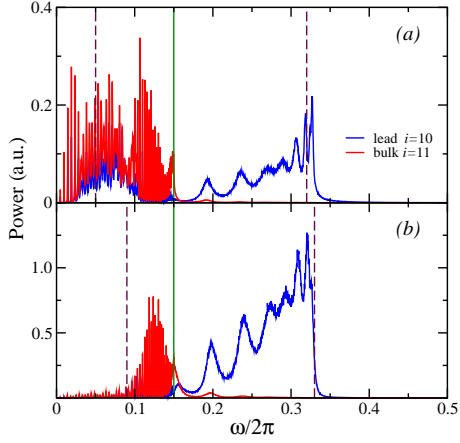}}
\caption{(Color online) Same as described in the caption of Fig.~\ref{fig:nueve}, but for $\gamma=0.25$ (a) $\Tl<\Tr$ and (b) $\Tl>\Tr$.}
\label{fig:once}
\end{figure}

The spectra corresponding to both temperature-bias instances are presented in Fig.~\ref{fig:once}; the phonon bands are the same as those in Fig.~\ref{fig:nueve}. Deviations from the frequency limits obtained from the effective phonon approach are less than those observed in Fig.~\ref{fig:siete}. This is because the system is in a temperature regime wherein the dynamics is closer to the harmonic limit, which increases the agreement with the effective phonon approach results. For the reverse-bias configuration the contribution of the phonons in the overlapping frequency range of both spectra is diminished relative to that obtained for the forward-bias one. This correlates well with the overwhelming contribution of $\jis$ to $J_i$ in the left lead region. All these features are consistent with a high rectification figure associated with $\gc=0.25$ reported in Fig.~\ref{fig:dos}(b).

To end this section we briefly mention the behavior of $r$ for $\gamma=10$. In this extreme case there is an overlap of the power spectra corresponding to the $J_-$ configuration in the high-frequency region, which increases the ensuing heat flux, thus reducing the rectification value. Furthermore, in this case the NNN harmonic interactions become more significant than the anharmonic ones associated with the leads, which drive the system close to the harmonic regime where it is well known that rectification does not exist~\cite{Pereira11b}.

%%%%%%%%%%%%%%%%%%%%%%%%%%%%%%%%FIG. 11%%%%%%%%%%%%%%%%%%%%%%%%%%%%%%%%%%%%%%%
%%%%%%%%%%%%%%%%%%%%%%%%%%%%%%%%FIG. 11%%%%%%%%%%%%%%%%%%%%%%%%%%%%%%%%%%%%%%%

\subsection{Rectification dependence on model parameters}

\begin{figure}
\centerline{\includegraphics*[width=80mm]{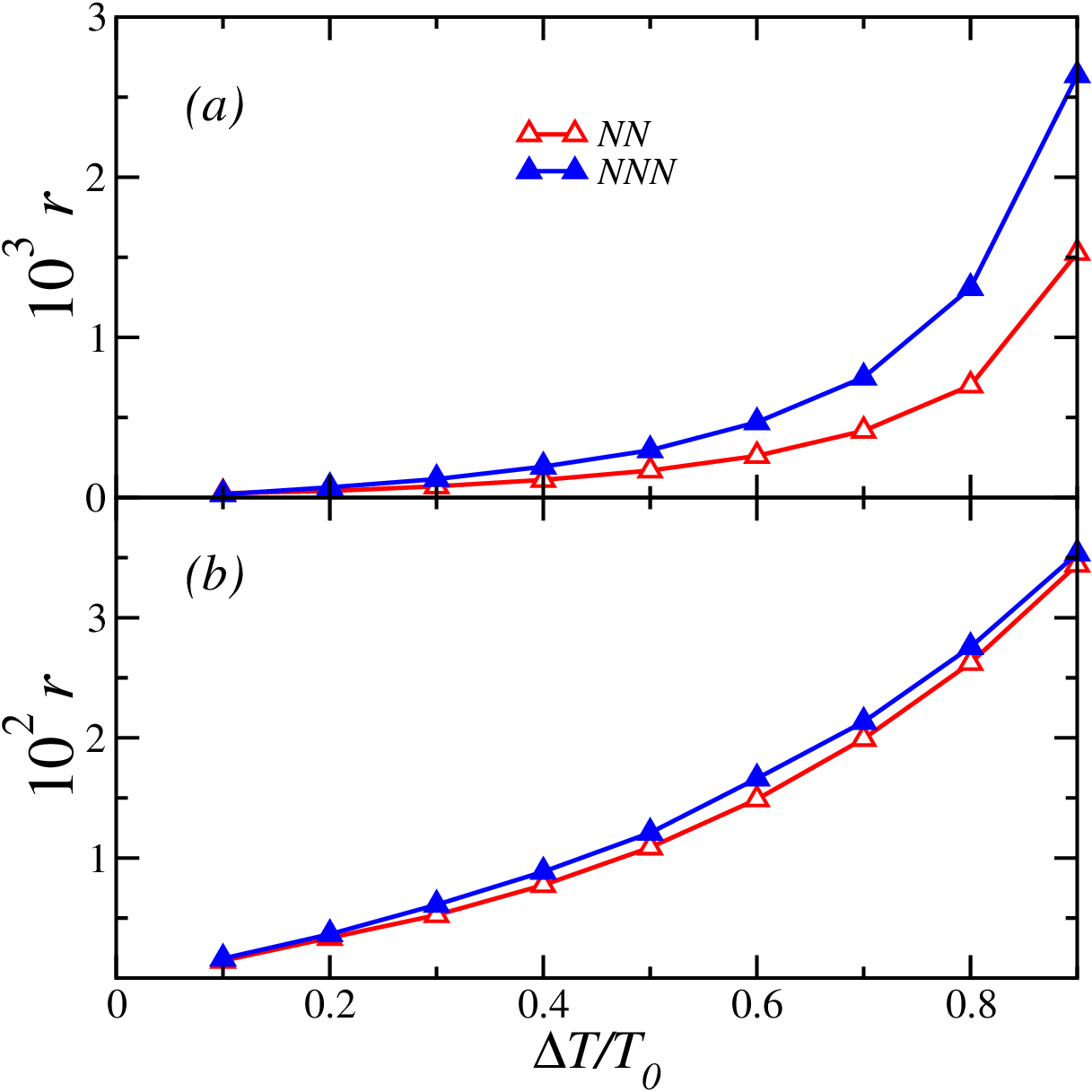}}
\caption{(Color online) (a) Dependence of thermal rectification $r$ on temperature difference $\DT/\To$ for $\gc=0.6$, $\To=5$, and $N=84$. Full symbols correspond to the lattice with both NN and NNN interactions and void ones to the lattice with only NN interactions. Same values of $\ml$, $\mr$, $\nl$, $\nr$, $\al$, and $\ar$ as in Fig.~\ref{fig:dos}(a). (b) Same as in panel (a), but now for $\gc=0.25$, and $\To=0.1$. Continuous lines are a guide to the eye.}
\label{fig:doce}
\end{figure}

In the following we analyze the dependence of the rectification efficiency on other parameters that determine the behavior of the system. In Fig.~\ref{fig:doce} we plot $r$ as a function of the temperature difference $\Delta T/\To$ in the absence and presence of NNN interactions in the high an low temperature regimes for the corresponding $\gc$ values. In both instances the behavior is the same in the $\Delta T/\To\rightarrow0$ limit: rectification steadily decreases as the linear response regime is attained. However, in the opposite limit the lattice with NNN interactions outperforms the one with only NN ones at an increasing rate as the temperature difference increases. Now, the most important difference is that, at the low $\To$ value, the rectification efficiency of both types of lattices is almost the same in the whole $\Delta T/\To$ value range studied.

%%%%%%%%%%%%%%%%%%%%%%%%%%%%%%%%FIG. 12%%%%%%%%%%%%%%%%%%%%%%%%%%%%%%%%%%%%%%%
%%%%%%%%%%%%%%%%%%%%%%%%%%%%%%%%FIG. 12%%%%%%%%%%%%%%%%%%%%%%%%%%%%%%%%%%%%%%%

\begin{figure}
\centerline{\includegraphics*[width=80mm]{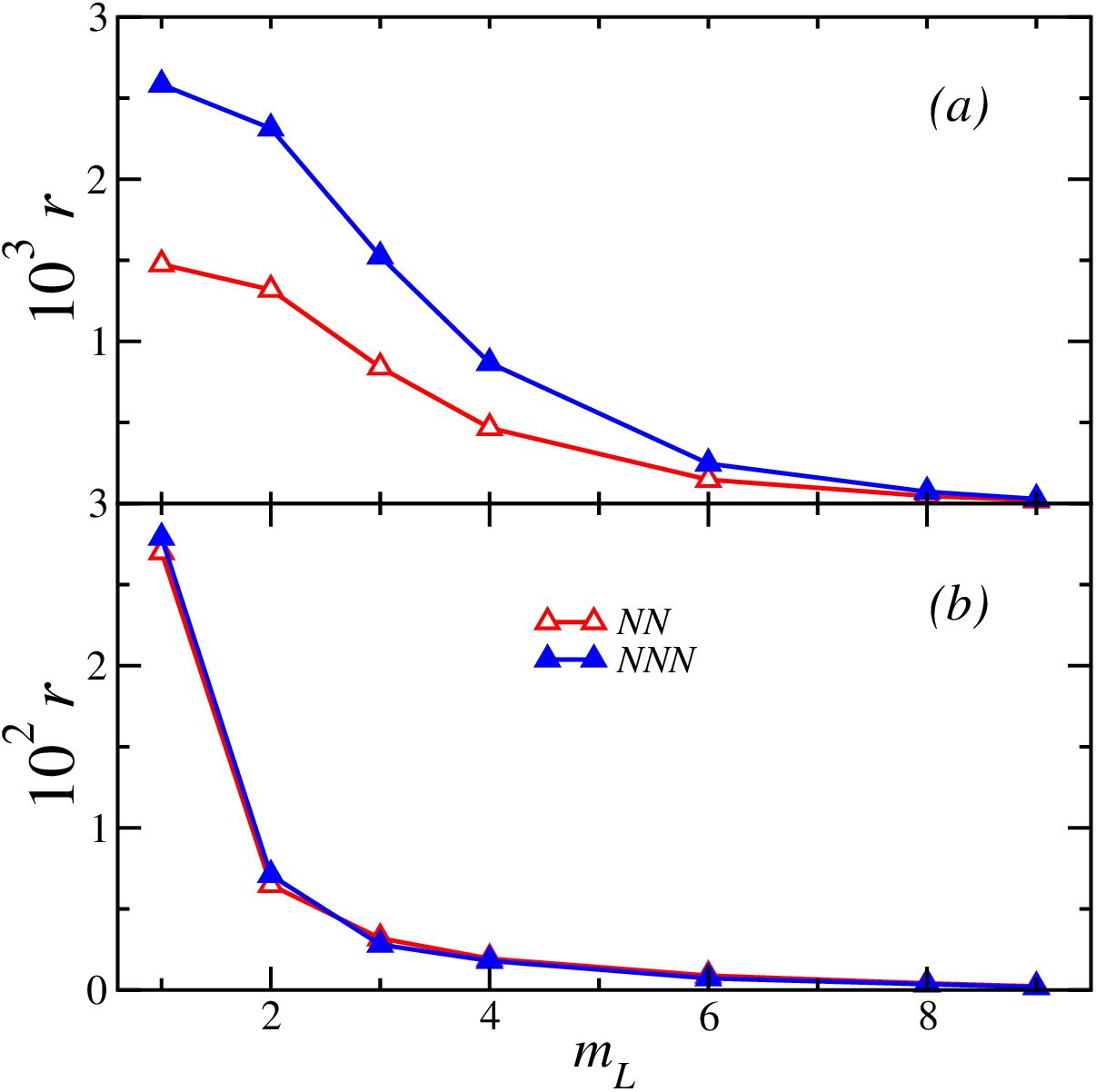}}
\caption{(Color online) (a) Dependence of thermal rectification $r$ on mass-lead value $\ml$ for $\gc=0.6$, $\To=5$, $\DT=9$, and $N=84$ with $\mc=(\ml+\mr)/2$. Full symbols correspond to the lattice with both NN and NNN interactions and void ones to the lattice only with NN interactions. Same values of $\mr$, $\nl$, $\nr$, $\al$, and $\ar$ as in Fig.~\ref{fig:dos}(a). (b) Same as in panel (a), but now for $\gc=0.25$, $\To=0.1$, and $\DT=0.16$. Continuous lines are a guide to the eye.}
\label{fig:trece}
\end{figure}

As for the dependence of $r$ on the mass $\ml$ of the oscillators in the left lead, for the high-temperature case the high $r$ value obtained for $\ml\ll\mr$ and depicted in Fig.~\ref{fig:trece}(a) is further increased by the NNN interactions. It decays in the same way as the case for the NN interactions, and thus moderate rectification values are still obtained for larger $\ml$ values. In the low-temperature case the behavior of $r$ seems to be independent of the existence of NNN interactions and the decay in $r$ as $\ml$ increases is stronger than the one obtained for high $\To$ values, as can be readily appreciated in Fig.~\ref{fig:trece}(b).

%%%%%%%%%%%%%%%%%%%%%%%%%%%%%%%%FIG. 13%%%%%%%%%%%%%%%%%%%%%%%%%%%%%%%%%%%%%%
%%%%%%%%%%%%%%%%%%%%%%%%%%%%%%%%FIG. 13%%%%%%%%%%%%%%%%%%%%%%%%%%%%%%%%%%%%%%

\begin{figure}\centering
\centerline{\includegraphics*[width=90mm,angle=-90]{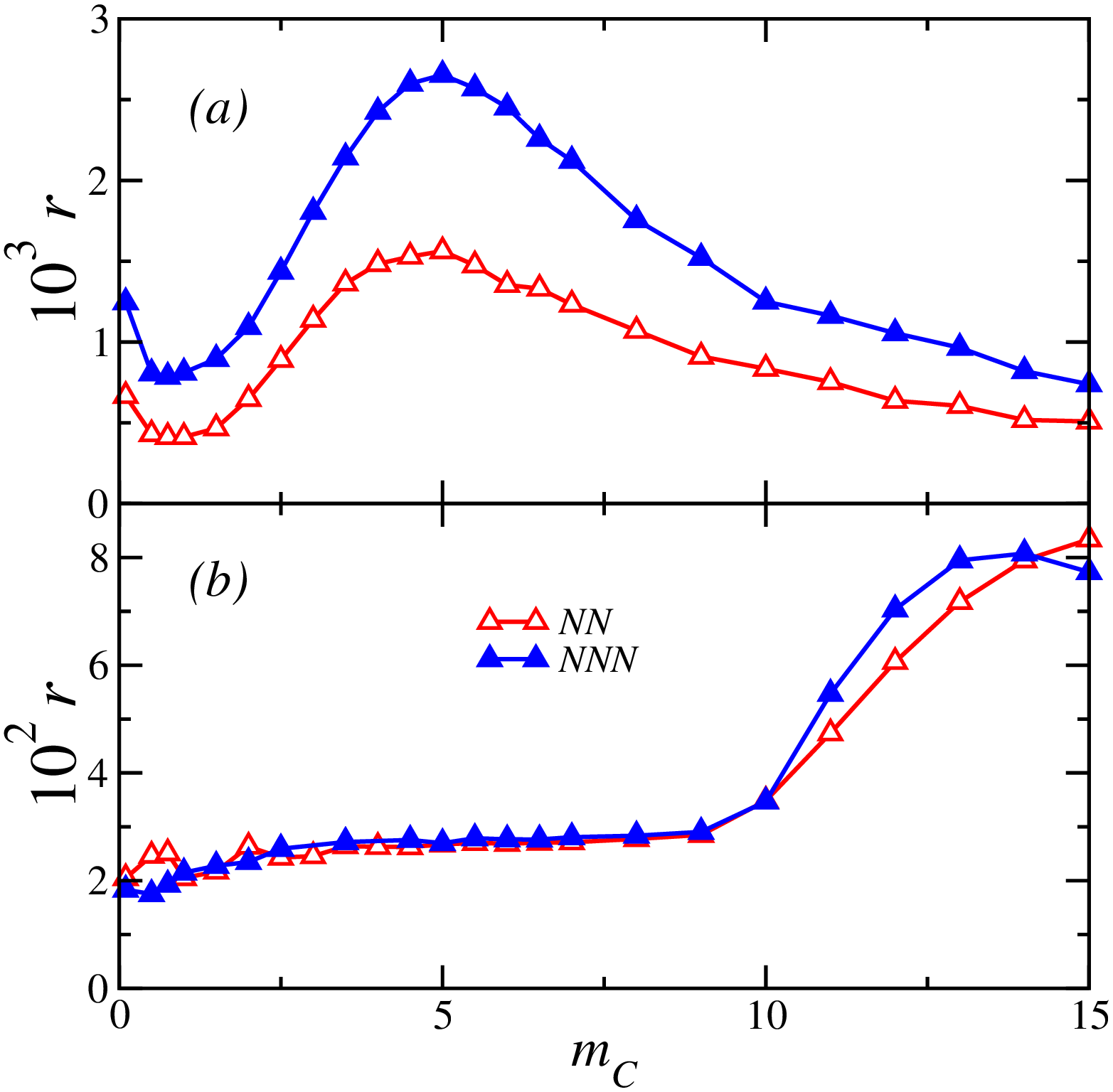}}
\caption{(Color online) (a) Dependence of thermal rectification $r$ on oscillator mass $\mc$ in the ballistic channel for $\gc=0.6$, $\To=5$, $\DT=9$, and $N=84$. Full symbols correspond to the lattice with both NN and NNN interactions and void ones to the lattice with only NN interactions. Same values of $\nl$, $\nr$, $\al$, and $\ar$ as in Fig.~\ref{fig:dos}(a). (b) Same as in panel (a), but now for $\gc=0.25$, $\To=0.1$, and $\DT=0.16$. Continuous lines are a guide to the eye.}
\label{fig:catorce}
\end{figure}

The behavior of $r$ as a function of the mass of the ballistic spacer $\mc$ is presented in Fig.~\ref{fig:catorce}(a) for the high-temperature case. The qualitative behavior is basically the same with and without NNN interactions, with a maximum rectification efficiency at the $\mc\simeq4.5$ already employed. The behavior of $r$ for the low-temperature regime depicted in Fig.~\ref{fig:catorce}(b) is, however, very different: for $\mc\lesssim10$ rectification remains almost constant and in the opposite regime presents a gradual growth until a maximum is attained at $\mc\simeq15$. And again, the best efficiency value $r$ is obtained in the presence of NNN interactions.

%%%%%%%%%%%%%%%%%%%%%%%%%%%%%%%%FIG. 14%%%%%%%%%%%%%%%%%%%%%%%%%%%%%%%%%%%%%%
%%%%%%%%%%%%%%%%%%%%%%%%%%%%%%%%FIG. 14%%%%%%%%%%%%%%%%%%%%%%%%%%%%%%%%%%%%%%

Further increments in rectification could be achieved for specific values of the aforementioned parameters in particular temperature intervals, i.e., with a large oscillator mass $\mc\simeq15$ in the ballistic channel for low average temperatures and with the $\mc$ value so far employed in the high average temperature range, both for $\DT/\To\gtrsim0.8$ values. The mass-lead value $\ml$ has to be low to maintain the asymmetry of the system. The oscillator number $n_{_{L,R}}$ in the leads also has to be low to confine the asymmetric local heat flow response corresponding to the NNN contribution close to the system boundaries, thus increasing the ensuing rectification value. In fact, $n_{_{L,R}}=8$ renders a marginally better rectification figure for the considered system sizes, with the rest of the parameters being the same as those reported in Fig.~\ref{fig:dos} caption, but $r$ always diminishes for $n_{_{L,R}}\ge10$ values.

%%%%%%%%%%%%%%%%%%%%%%%%%%%%%%%%%%%%%%%%%%%%%%%%%%%%%%%%%%
\section{concluding remarks\label{sec:Disc}}
%%%%%%%%%%%%%%%%%%%%%%%%%%%%%%%%%%%%%%%%%%%%%%%%%%%%%%%%%%

In summary, we have performed the study of the rectification properties of a NNN coupled anharmonic mass-graded lattice with a ballistic spacer. Its performance is enhanced in comparison to the previously considered system with only NN interactions for the value range of the considered parameters and the rectification efficiency is maximized for a certain value of the coupling constant of the NNN interactions. It is higher to that obtained for the linear mass-graded lattice previously studied~\cite{Romero17} when the device operates in the high-temperature regime and much lower in the opposite one. In both instances the temperature difference is of the same order as the average temperature; therefore, the size-independent thermal gradient affords a significant rectification figure. However, for a low average temperature the TR value is almost independent of the NNN coupling value in a wider range of the latter. Thus, although the rectification is an order of magnitude lower than that for high average temperature, it is more stable against details of the NNN interactions. This property enhances its usefulness considering future applications. By means of the spectral analysis we have determined that the rectification properties stem from a surface-like behavior of the flux associated with the NNN interactions ---which is dominated by the contribution of high-frequency phonons--- at the leftmost, lighter lead, in a manner similar to what happens in the linear mass-graded lattice. Thus, it would be of interest to explore how this interface behavior is affected with other asymmetries applied on the end leads, such as defect, geometry, chemical functionalization, substrate couplings, mechanical strains, etc. Finally, we recall that recently a 1D graded rotor lattice with only NN interactions has been proposed as model wherewith TR is enhanced in the large system-size limit~\cite{You20} by means of the suppression of heat flux by local nonlinear modes stimulated at the end of the lattice coupled to the hot reservoir. It is then natural to enquire how NNN interactions may affect the nonlinear modes and thus TR. The herein studied model allows to explore in a systematic way the influence of interactions of gradually increasing range (third, fourth neighbors and so on), and is well suited to address the above posed question.

\begin{acknowledgments}
M.~R.~B. thanks Consejo Nacional de Ciencia y Tecnolog\'\i a (CONACYT) Mexico for financial support. J.~I.~A.~D. thanks ``Programa Institucional de Formaci\'on de Investigadores" I.P.N. M\'exico for financial support. M.~R.~B. also thanks Stephane Duran and Joaqu\'\i n Garc\'\i a-Aguila for useful comments and discussions.
\end{acknowledgments}

%%%%%%%%%%%%%%%%%%%%%%%%%%%%%%%%%%%%%%%%%%%%%%%%%%%%%%%%%%%%%
\bibliographystyle{prsty} % Older version of RevTeX4
%\bibliographystyle{apsrev4-1} % Current RevTeX4
%\bibliography{Bibliography}

\end{document}